# Better Quantum Seal Schemes based on Trapdoor Claw-Free Functions

Xiao-gang CHENG[*1,2], Ren GUO[3]

1. College of Computer Science and Technology, Huaqiao University, Xiamen 361021, China

2. Xiamen Key Laboratory of Data Security and Blockchain Technology, Xiamen 361021, China

3. College of Business Administration, Huaqiao University, Quanzhou 362021, China

*Corresponding author: cxg@hqu.edu.cn

**Abstract:** Seal in classical information is simply impossible. Since classical information can be easily copied any number of times. Based on quantum information, esp. quantum unclonable theorem, quantum seal maybe constructed perfectly. But it is shown that perfect quantum seal is impossible, and the success probability is bounded. In this paper, we show how to exceed the optimal bound by using the TCF (Trapdoor Claw Free) functions, which can be constructed based on LWE assumption. Hence it is post-quantum secure.

## 1. Introduction

Seal is a kind of cryptography scheme [1], by which Alice can send a sealed information to another party Bob. Bob should only open the seal and get the hidden information when Alice order him to do so or some predefined condition is satisfied like a predefined date/time. And Alice has a way to check the honesty of Bob by asking him to send back the seal before the open order or the predefined date, then Alice can check if the seal has been opened or not.

In classical digital information technology, this is simply impossible. Since any information can be copied any number of times classically. Bob can simply copy the seal and open it, then send back the original copy without being detected. In quantum information, there is a fundamental no-cloning theorem, to prevent copying of an unknown quantum state. Hence a natural idea is to construct cryptographic seal scheme based on quantum information, esp. the no-cloning theorem [1,2].

There are some constructions and discussions on quantum seal [3-11], since it is introduced by Pasquinucci in [1] in 2003. But it is shown that perfect quantum seal is not possible [3,4]. There is fundamental tradeoff between Bob's successful reading probability and the probability that Alice can find the cheating behavior. I.e., if Bob can open and read the sealed information with high probability, then Alice's probability of finding Bob's cheating will be low. High probability for both is not possible even in quantum information [7,10]. In this paper, we show how this optimal bound can be exceeded by a construction based on TCF function.

Quantum seal are also related other quantum cryptographic schemes such as quantum bit commitment [11,12] and quantum OTM (One-Time Memory) [13]. And quantum seal is weaker than quantum OTM [14] and stronger than quantum bit commitment. I.e., quantum bit commitment can be built from quantum seal, while a nearly perfect quantum seal can be built from quantum OTM. It also means that a no-go bound for quantum bit commitment also applies to quantum seal, but a no-go result for OTM does not necessarily apply to quantum seal.

It was shown in [15] that random bits generated by quantum device can be verified by TCF functions. In [16], TCF functions was used for a classical computer to verify the computation result of a quantum computer, which will be a common scenario in quantum cloud and delegation computation in the future. Claw just means two pre-images, which are map to the same image under the TCF map. With a trapdoor, a claw can be easily found. I.e., two different pre-images $x_1$ and $x_2$ that are map to the same image under the TCF function, $y = f(x_1) = f(x_2)$. On the other hand, without the trapdoor, it is hard to find such a claw even for a quantum computer.

In this paper, we show how the optimal bound in [7,10] can be exceeded by introducing TCF to the construction of quantum seal. Note that our result does not break the no-go result in [7,10], since TCF based construction rely on computational assumptions, such as the learning with errors assumption for quantum computer [17,18].

In [7,10], it is shown that the optimal bound of Alice's honesty checking probability is 50% when Bob's successful reading probability is one hundred percent. I.e., when Bob breaks the seal and steals the information inside it, he can escape Alice's checking half of the times at least. We show that this bound can be improved to 85% with our TCF-based construction. We also show that same 50% probability can be achieved while Bob is only required to return classical message instead of quantum, hence it is more convenient and efficient in practice.

In section 2, some preliminary knowledges are given. Our TCF-based quantum seal construction is given in section 3. In section 4, the security of our quantum seal scheme is analyzed and compared with previous optimal bound. And we conclude in section 5.

## 2. Preliminaries

Quantum seal is a cryptographic protocol by which Alice can send a sealed information to Bob. And Bob should only open the seal and get the information when Alice gives the open order or some other conditions are satisfied. It works as follows:

    1. Alice send a quantum state to Bob, together with some description about how to measure the quantum state to get the classical message encoded in the quantum state. But ask Bob to extract the message only when some conditions are satisfied, e.g., a pre-defined date/time, or some jobs have been done by Bob.

    2. After get the quantum state and the measurement instructions from Alice, Bob can measure the quantum state and get the message with high probability ($P_{read}$) any time.

    3. Alice can check the honesty of Bob by the following protocol. At any time

before the afore-mentioned conditions are satisfied. Alice can ask Bob to send back the quantum state, and check if Bob has broken the seal or not. With probability $P_{check}$, Alice can detect Bob's cheating if Bob is dishonest.

It is shown that it is impossible for both $P_{read}$ and $P_{check}$ to be high simultaneously [7,10]. Especially when $P_{read} = 1$, namely Bob can open and read the sealed message with one hundred percent probability, then Bob's chance of escaping Alice's honesty checking probability is 50% at least when he is dishonest.

TCF (Trapdoor claw-free) function $f$ is a cryptographic map with the following properties:

For any $x$, it is easy to calculate the map

$$y = f(x)$$

Given $y$, it is hard to compute the pre-image $x$. But with a trapdoor, not only the pre-image of $y$ can be found, another pre-image x' can also be found efficiently, i.e.,

$$x \neq x', y = f(x) = f(x')$$

Such a pair $(x, x')$ is called a claw.

It is shown that TCF functions can be constructed [15-17] from the well-known quantum secure assumptions LWE (Learning with Errors) assumption [18].

## 3. Our Construction

Our TCF-based quantum seal scheme works as follows:

1. Alice prepares the following quantum state:

$$(|x_1\rangle + |x_2\rangle)|y\rangle$$

Where $x_1$ and $x_2$ are the pre-image of $y$, i.e.,

$$f(x_1) = f(x_2) = y$$

Where $f$ is a TCF function as we mentioned above.

Then Alice sends the superposition quantum state

$$|x_1\rangle + |x_2\rangle$$

to Bob. And the corresponding $y$ is the secret inside the quantum seal.

2. If Bob wants to open the seal and get the secret in it, he can measure the quantum state he received from Alice by the computational basis $\{|0\rangle, |1\rangle\}$. And get one $x \in \{x_1, x_2\}$, then by applying the TCF function $f$, Bob will get

$$y' = f(x)$$

Obviously $y'$ and $y$ should be the same if both parties follow the protocol.

3. If Alice wants the secret back, then Bob has to prove that he has not try to steal the secret inside the seal. There are two strategies:

1) Quantum information back:

Bob has to send back the superposition state

$$\frac{1}{\sqrt{2}}|x_1\rangle + |x_2\rangle$$

back to Alice. If Bob is honest, he can just simply send the original quantum superposition state back. If Bob has stolen the secret, the result quantum state would be either $|x_1\rangle$ or $|x_2\rangle$ instead of the superposition state. Then Alice would have a high probability to find the cheating behavior of Bob.

The trace distance between $|x_2\rangle$ (or $|x_1\rangle$ ) and $1/\sqrt{2}(|x_1\rangle + |x_2\rangle)$ is:

$$\delta(|x_2\rangle, |x_1\rangle + |x_2\rangle) = \sqrt{1 - \frac{1}{2}|\langle x_2|x_1 + x_2\rangle|^2} = \frac{\sqrt{2}}{2}$$

Hence the probability that Bob's cheating behavior will be detected by Alice is

$$\frac{1}{2} + \frac{1}{2} * \delta \approx 0.85$$

at least.

If instead of sending the result state $|x_1\rangle$ (or $|x_2\rangle$ ), Bob could also send a random quantum state $|r\rangle$ to Alice. But this can only increase the probability of being caught cheating. Since now the trace distance is:

$$\delta(|r\rangle, |x_1\rangle + |x_2\rangle) = \sqrt{1 - \frac{1}{2}|\langle r|x_1 + x_2\rangle|^2} \approx 1$$

Since if $|r\rangle$ is a pure state in computational basis, i.e., $|r_1 ... r_n\rangle$, then the probability of being same with $|x_1\rangle$ or $|x_2\rangle$ is $\frac{2}{2^n}$. Hence with probability $1 - \frac{2}{2^n}$ , the trace distance is 1. And Alice's detecting probability is:

$$\frac{1}{2} + \frac{1}{2} * \delta \approx 1$$

If bob choose to send a superposition state, he cannot be more successful. Since Bob only knows one value in $\{x_1, x_2\}$, and is ignorant of the other one.

2) Classical information back

Instead of asking Bob sending back the quantum superposition state, which is hard to implement in practice. Alice can also ask Bob to return a classical bit string $d$, which should satisfy the following relation:

$$d \cdot (x_1 \oplus x_2) = 0$$

If Bob has not measured the quantum superposition state $|x_1\rangle + |x_2\rangle$, he can easily get the bit string $d$ as the following. Perform a Hadamard transform on each qubit of the superposition state $|x_1\rangle + |x_2\rangle$, then the result state is:

$$\frac{1}{\sqrt{2^n}} \sum_d ((-1)^{d \cdot x_1} + (-1)^{d \cdot x_2})|d\rangle$$

By measure in the computational basis, Bob can easily get a random bit string d, which satisfy the relation mentioned above.

Otherwise, if Bob have measured the superposition state and try to open the quantum seal, he will not be able to produce such a bit string.

Based on the security of TCF functions, Bob would not be able to produce this bit string if he has already measured the superposition quantum state by the computational basis and stolen the secret. And Bob also cannot get both $x_1$ and $x_2$, so the best strategy of Bob is to choose a random bit string $d$. Obviously, a random bit string only has 50 percent chance for satisfying the equation. And 50 percent chance of being caught for the cheating behavior.

Note that this 50% probability of catching the cheating behavior of Bob has already match the optimal bound in [7,10], when Bob's reading probability is 100%. And the 85% probability for the quantum back strategy has exceeded the optimal bound. But it does not break the no-go result, since our construction is based on TCF functions, which are currently constructed from computational assumption LWE. I.e., our construction is computationally secure other than informatically secure.

## 4. Security analysis and efficiency comparison

To prove the optimal bound of quantum seal scheme, the strategy in [10] is to carry out the following POVM measurement $\{F_i\}_{i \in [M]}$:

$$F_i = \sum_{j \in [M_i]} \varepsilon_{i,j}$$

Where $\varepsilon_{ij}$ are the elements of Alice's recommended POVM for Bob to extract information from the quantum seal. It is shown in [10] that by using this proof strategy, if Alice wants Bob to read the message with high probability, then with high probability that she will not be able to detect Bob's cheating behavior. Especially, the optimal probability $P_{check}$ is only 50%, when we want that Bob's successful reading probability $P_{read}$ is 100%.

Now we show that this strategy fails in our construction. In our TCF based quantum seal construction, there are two messages that are equally good, namely $x_1$ and $x_2$. Since the images of these two messages under the TCF map are the same, i.e., $f(x_1) = f(x_2) = y$, which is the secret hidden in the quantum seal.

Alice's suggested measurement is by computational basis. Hence the message is a random n-bit long number $i$, $i \in [0, 2^n - 1]$. if Alice quantum message is just $|x_1\rangle$, then after the measurement, Bob will get the message $x_1$ with 100 percent probability. And Bob just sends back the state after measurement, which is still $|x_1\rangle$, then his cheating behavior will not be detected since the state is not changed. Basically, this the essential idea in [10], to prove the optimal bound of quantum seal.

But now in our TCF based quantum seal construction, there are two equally good messages $x_1$ and $x_2$, either one can be used for extracting the secret $y$. Hence the aforementioned strategy failed. Since after computational basis measurement, the result state will still have significantly difference with the original superposition state, as calculated in the previous section.

We could extend our construction as following to further increase the optimal probability bound of quantum seal, but with classical symmetric encryption and more overheads. Instead of sending a superposition state of two random message, Alice could send a superposition state of three random message:

$$\frac{1}{\sqrt{3}}(|x_1\rangle + |x_2\rangle + |x_3\rangle)$$

to Bob. The secret $y$ is also encrypted by the three random messages as keys:

$$C_1 = ENC_{x_1}(y), C_2 = ENC_{x_2}(y), C_3 = ENC_{x_3}(y)$$

Alice then sends the quantum state and the three cyphertexts to Bob. Obviously, Bob can get one of the messages by measuring in the computational basis, and then decipher one the cyphertext and get the secret $y$.

And now the probability of Bob's cheating behavior being caught will increased. Since the trace distance will be increased from $\frac{\sqrt{2}}{2}$ to:

$$\sqrt{1 - \frac{1}{3}|\langle x_2|x_1 + x_2 + x_3\rangle|^2} = \frac{\sqrt{6}}{3}$$

And the probability $P_{check}$ will be increased from 0.85 to:

$$\frac{1}{2} + \frac{1}{2} * \frac{\sqrt{6}}{3} \approx 0.9$$

Similarly, if 4 random bit strings are superposed, then the probability $P_{check}$ will be increased to about 0.93. And with 5 random bit strings, the probability is about 0.95.

In general, if $n$ random bit strings are superposed, the probability $P_{check}$ will be:

$$\frac{1}{2} + \frac{1}{2} * \sqrt{1 - \frac{1}{n}}$$

Which is plotted in Fig 1. So, when $n$ approach infinity, both the probability $P\_check$ and $P_{read}$ will be 1, i.e., we get a perfect quantum seal scheme.

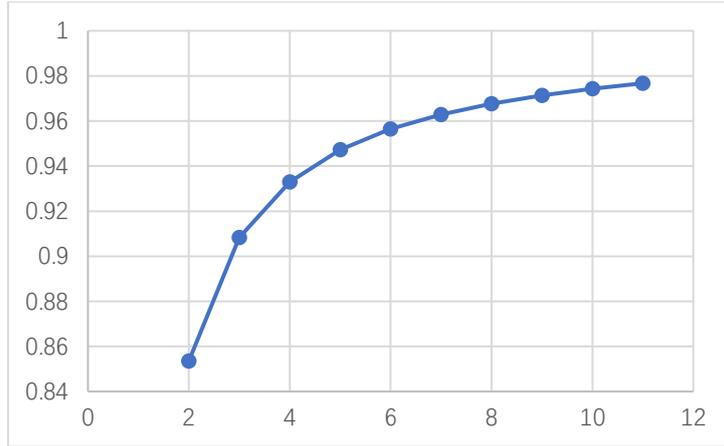

Fig 1. Alice's probability of detecting Bob's cheating behavior versus number of superposed bit strings

It is obvious that with more superposed bit strings, more secure is the quantum seal scheme. But more overheads are needed, since $n$ ciphertexts should be used if there are $n$ number of superposed keys. And they should be sent to Bob together with the superposition quantum state. Then by one of the keys, Bob can decrypt one of the ciphertexts and get the information.

## 5. Conclusions

In this paper, we present a way to exceed the optimal bound of quantum seal by using TCF functions. The optimal probability of Alice's successful detecting Bob's dishonesty $P_{check}$ is 50% when Bob's reading probability $P_{read}$ is 100%. In our TCF function-based construction, we improve $P_{check}$ to 85% when $P_{read}$ is 100%. We also show that a perfect quantum seal scheme can be constructed asymptotically with more overheads based on quantum superposition state and symmetric encryptions.

Note that the no-go result of quantum seal in [7,10] is not violated, since TCF functions are constructed based on LWE assumptions. Namely our quantum seal construction is computationally secure, instead of informatically secure.

There are some interesting future research directions. Firstly, we plan to further improve the quantum seal probability bound by using other quantumly secure assumptions. And even construct perfect quantum seal scheme without more overheads, which are computationally secure. Secondly, we plan to extend our construction to many other related cryptographic schemes such as OTM and bit commitment etc.

## Declarations

Funding. This work was supported by Social Science Foundation of Fujian Province, China (FJ2020B044, FJ2021B163).

Competing interests. The authors have no financial or proprietary interests in any material discussed in this article.

Data sharing not applicable to this article as no datasets were generated or analyzed during the current study.